\documentclass[traditabstract,onecolumn]{aa}
\usepackage{graphicx}
\usepackage{epstopdf}
\usepackage{amstext,amssymb,amsbsy,amsmath}
\usepackage{txfonts}
\usepackage{Natbib}
\usepackage{longtable}

\begin{document}

\title{1318 New Variable Stars in a 0.25 Square Degree Region of the Galactic Plane
\footnote{Table 2 and also the photometric data for the variable stars are only available in electronic form 
at the CDS via anonymous ftp to cdsarc.u-strasbg.fr (130.79.128.5) or via http://cdsweb.u- strasbg.fr/}
}

\titlerunning{1318 New Variable Stars}

\author {
V.R.~Miller 
\inst{1,2}
\and
M.D.~Albrow 
\inst{2}
\and
C.~Afonso
\inst{1}
\and
Th.~Henning
\inst{1}
}

\institute{
Max-Planck Instit\"{u}t f\"{u}r Astronomie, K\"{o}nigstuhl 17, Heidelberg, 69117, Germany
\email{vrmiller@live.com, afonso@mpia.de, henning@mpia.de}
\and
Department of Physics and Astronomy, Private Bag 4800, University of Canterbury, New Zealand \\
\email{Michael.Albrow@canterbury.ac.nz}
}

\date{Received  ; accepted  }

\abstract 
{We have conducted a deep photometric survey of 
 a 0.5$^\circ$ x 0.5$^\circ$ area of the Galactic Plane using the WFI instrument on the 2.2-m ESO telescope on La Silla, Chile. The dataset comprises a total of 267 R-band images,  204 from a 16 day observation run in 2005, supplemented by  63 images from a six week period in 2002. Our reduction employed the new numerical kernel difference image analysis method as implemented in the \textsc{pysis3} code and resulted in more than 500,000 lightcurves of stars down to a magnitude limit of R $\sim$ 24.5. A search for variable stars resulted in the detection of 1318 variables of different types. 1011 of these are eclipsing or contact binary stars. A number of the contact binaries have low mass-ratios and several of the detached binaries appear to have low-mass components. Three candidate contact binaries  have periods at the known cut off including two with periods lower than any previously published. Also identified are 3 possible pre-main sequence detached eclipsing binaries. }


\maketitle

\section{Introduction}			\label{sec:intro}
During the past decade, many photometric surveys have been  aimed at the detection of 
extrasolar planets via transits or microlensing. A byproduct of these programmes is the 
detection of large numbers of variable stars \citep{Wozniak2002, Wyrzykowski2004, 
Soszynski2008a, Soszynski2008b, Soszynski2009, Bayne2002, Albrow2001a, Weldrake2004, Weldrake2007d}.

In contrast to previous surveys, which have generally been targeted at the Galactic Bulge, 
Magellanic Clouds or globular clusters, this survey is of a 0.25 square degree 
region of the Galactic Plane, centred on galactic coordinates (330.94,-2.28). The majority
of stars in this field are thought to be associated with the Norma Spiral Arm.
In this paper we  present
a comprehensive catalogue of variable stars that we have detected, most of which are binary stars.

\section{Observations}			\label{sec:obs}
All observations for this project were obtained using the WFI camera on the ESO 2.2m telescope at La Silla, Chile.
The initial images were taken during 2002 as part of a microlensing followup study by the PLANET collaboration \citep{Albrow1998}. A  follow-up observation run was made in 2005 in service mode using MPIA time (P.I. C. Afonso).

The WFI camera is a 4 by 2 mosaic of EEV CCD44 chips each containing 2k by 4k 15-$\mu$ pixels. The pixel scale is 0.238 arcsec/pixel. The total dataset (including the pilot data from 2002) is comprised of 267 R-band images, though a few were bad enough to be removed from the dataset prior to processing. For the most part the stellar images have a full width at half maximum of between 5 and 8 pixels. Due to inclement weather and other considerations, only half of the anticipated number of images was obtained during the 2005 observing run. A summary of the observations is given  in Table~\ref{tab:totobs}.

\section{Reduction}			\label{sec:red}

Bias-subtraction and flat-fielding of the images was performed using the \textsc{esowfi} and \textsc{mscred} packages in \textsc{iraf}. After the mosaic images had been pre-processed they were divided into individual FITS files for each CCD chip that were further sectioned into eight 1k by 1k sub-images, resulting in 64 file sets (eight sections of eight CCD chips).

Difference imaging was carried out on each of the 64 image sets using \textsc{pysis3} \citep{Albrow2009} .
This code employs an optimal numerical convolution kernel \citep{Bramich2008} to match a photometric reference template to each target image in turn. This pixel representation of the kernel is the major point of difference with respect to the \textsc{isis} code \citep{Alard1998}, which uses a sum of Gaussian functions multiplied by polynomials to define the kernel. For the current application, each pixel in the kernel was allowed to have a linear spatial variation across each 1k by 1k  subimage.

Prior to the difference-imaging process, the best-seeing image was chosen as the astrometric template, and all other images were aligned to the nearest pixel of this template using integral-pixel XY shifts. A feature of our reduction method is that complete registration is not required, meaning that image interpolation can be avoided.

After several tests, our photometric reference frame was chosen to be a stack of three of the best-seeing images in the dataset. All these images had the best seeing over all eight chips and were all from the 2005 dataset. 

\textsc{pysis3} was developed for reduction of microlensing events and as such is designed to obtain a single lightcurve of the target star. For this project the code was modified to cycle through a list of coordinates of star positions. Optimal PSF-fitting photometry of all identified stars was carried out as follows. First, the PSF of 
the reference image was computed using a combination of relatively isolated bright stars. This reference PSF was then convolved to each target image and  an optimal fit made at the coordinates of each star. 
The reduction of the dataset was accomplished using an IBM P575 supercomputer at the University of Canterbury's BlueFern facility, and resulted in difference-flux lightcurves for more than 500,000 stars. 

\begin{figure}
\resizebox{\hsize}{!}{\includegraphics{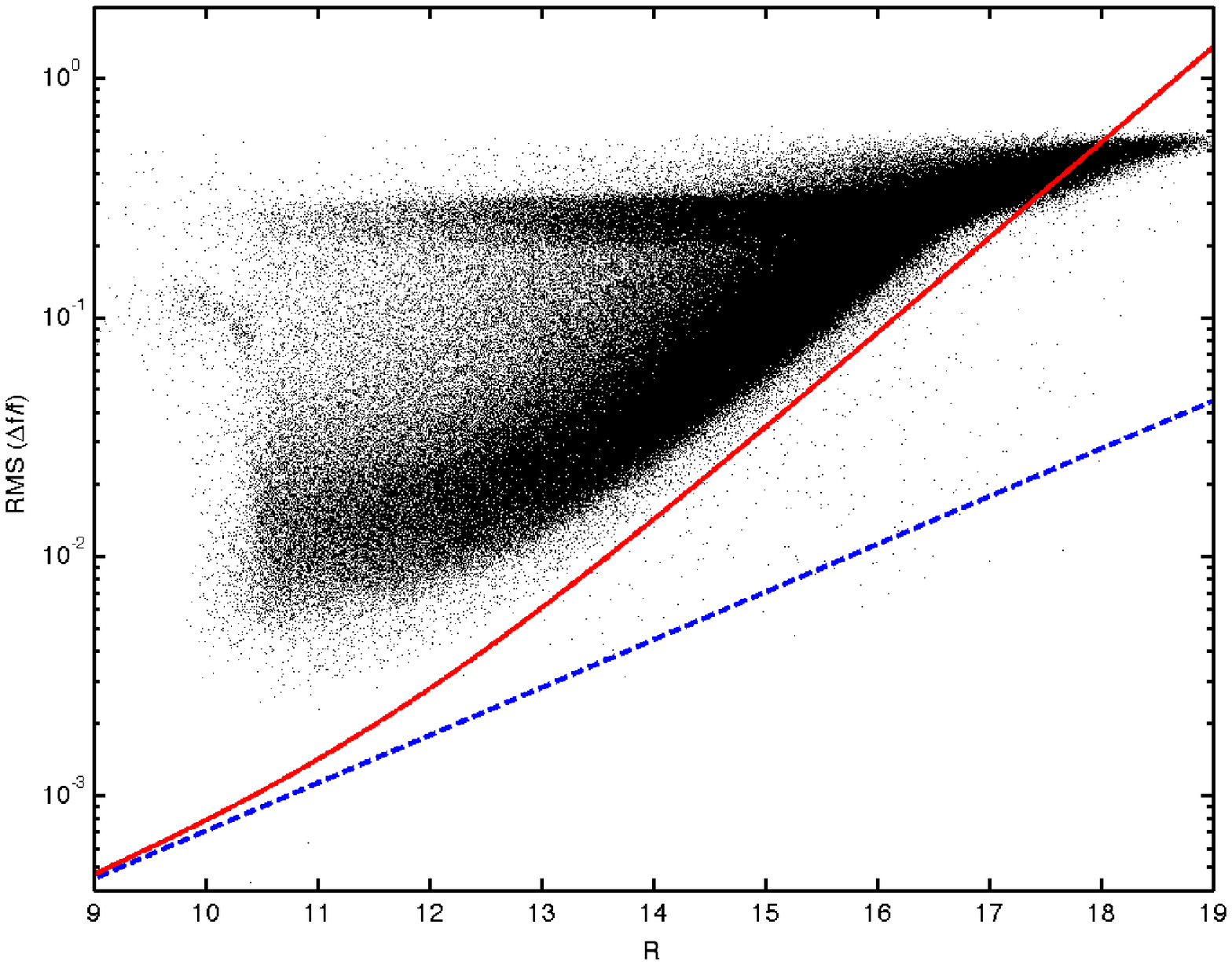}}
\caption{RMS scatter of ($\Delta f /f$) lightcurves as a function of R magnitude. Also plotted are Poisson noise with (upper line) and without (lower line) average sky and readout noise.}
\label{fig:RMSlcdt}
\end{figure}

The \textsc{daophot} \citep{Stetson1987} package in \textsc{pyraf} was used to find stellar positions and fluxes on the reference images.
Between seven and thirteen thousand stars were identified on each subimage using \textsc{daofind}. 
\textsc{allstar psf} photometry was performed to determine reference fluxes for all identified stars.
In  Figure~\ref{fig:RMSlcdt} we show the RMS scatter in the photometric measurements  as a function of R magnitude for the entire dataset.

\section{Calibration}			\label{sec:calib}

Using the astrometric capabilities of \textsc{skycat/gaia} \citep{Albrecht1997} and \textsc{aladin} \citep{Aladin} the x,y pixel coordinates of the reference frames have been transformed to J2000 sky coordinates defined using Digitized Sky Survey\footnote{This research has made use of the Digitized Sky Surveys, produced at the Space Telescope Science Institute under U.S. Government grant NAG W-2166. The images of these surveys are based on photographic data obtained using the Oschin Schmidt Telescope on Palomar Mountain and the UK Schmidt Telescope. The plates were processed into the present compressed digital form with the permission of these institutions.} (DSS) images and coordinates from  the United States Naval Observatory (USNO) catalogue~\citep{USNO2003}.

\begin{figure}
\resizebox{\hsize}{!}{\includegraphics{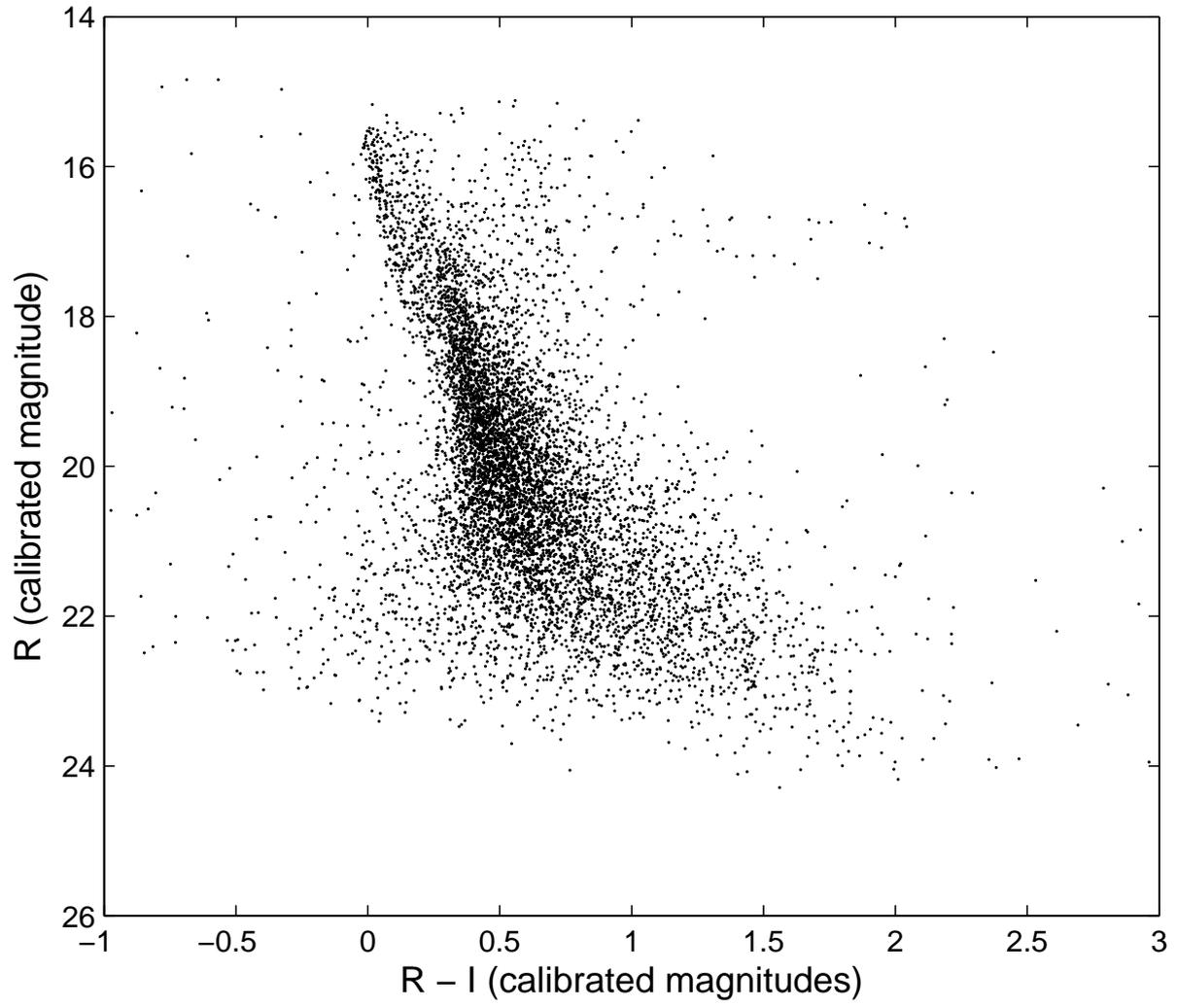}}
\caption[Colour-magnitude diagram]{Colour magnitude diagram of a subset of the field (1B).}
\label{fig:cmd}
\end{figure}

Similarly to the coordinate calibration, around 150 stars were identified in the reference frame from the USNO catalogue~\citep{USNO2003}. The USNO B1.0 R1 magnitudes were recorded and compared with those found on the reference template using \textsc{daophot}.
The average offset for the R magnitudes over the field as compared to the USNO B1.0 R1 magnitudes was used to correct our instrumental magnitudes. The corrections for B and I were made in similar fashion. We note that no colour-dependent terms have yet been applied and that our ``calibrated'' magnitudes are solely corrections to put them on the USNO B1.0 scale.
Figure~\ref{fig:cmd} shows a calibrated colour-magnitude diagram of a subset of the field.

\section{Variable search}		\label{sec:varsear}

The initial variable star search was made by using the Lomb-Scargle (LSA) period finding algorithm~\citep{Lomb1976}. 
The LSA was run on every lightcurve with trial periods distributed logarithmically between 0.01 and 10 days. The highest power was recorded with the attendant false alarm probability and period. The period range was chosen due to the limits on observations - there are only 16 consecutive days of observations that are well sampled.
Longer periods were considered but due to the uneven spacing of the data it is far less likely that a variable with period longer than 10 days (where we will see only a single cycle) will be detected with its correct period. However some longer period variables have been detected through their aliases.

In order to determine the appropriate threshold for variable star detections, simulated sinusoidal variables were inserted into a subset of the real data.  
The simulated variables had periods chosen randomly in the range 0.01 to 10 days and amplitudes in the range 0.05 to 0.5 mag, and were subjected to the same LSA procedure as the real data.
We found that 80\% of simulated variables had LSA false alarm probabilities (FAP) below 10$^{-8}$, with the FAP decreasing as the amplitude increases. Conservatively, we adopted 
10$^{-6}$ as our LSA FAP threshold.

\begin{figure}
\resizebox{\hsize}{!}{\includegraphics{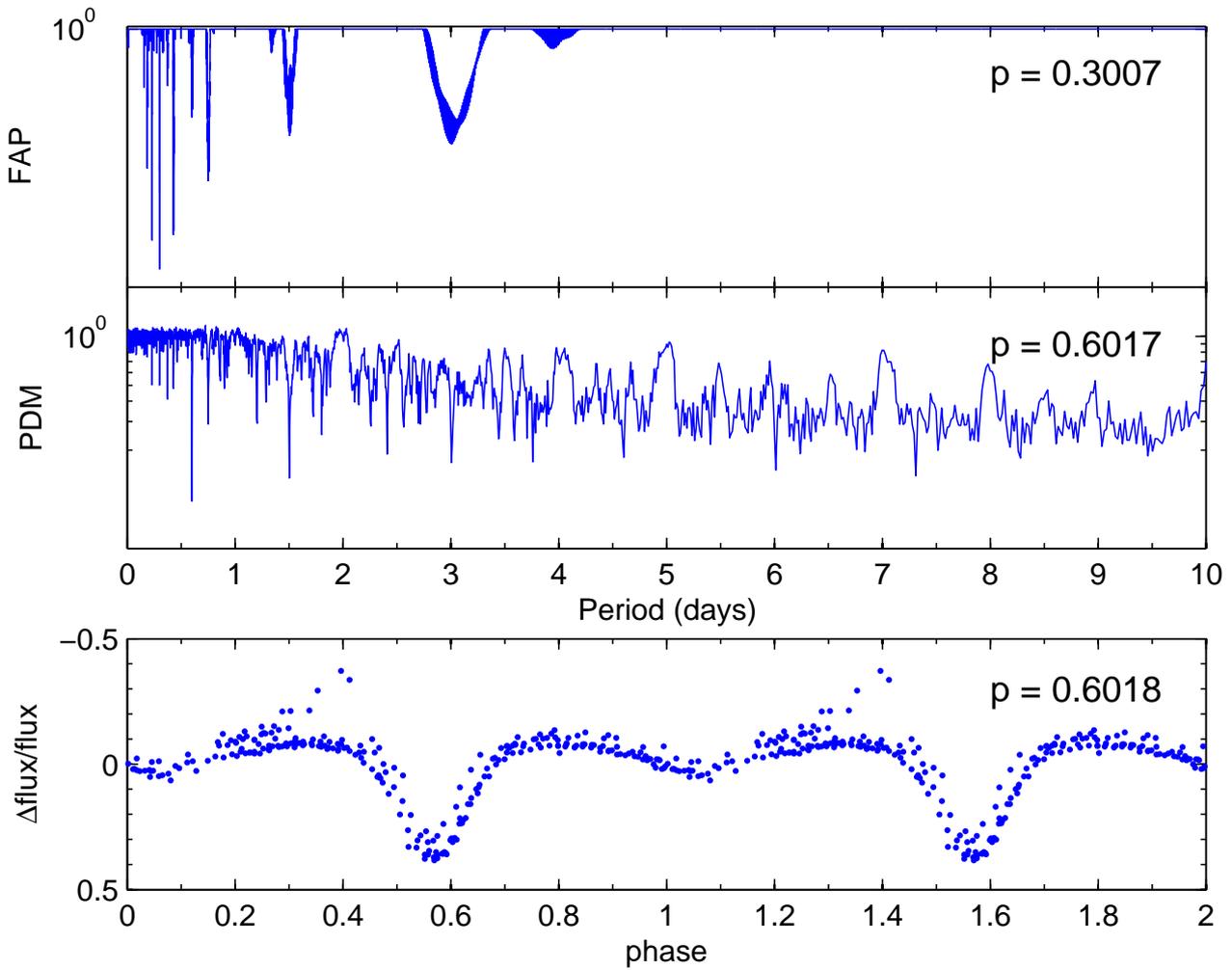}}
\caption{Periodograms for an eclipsing binary star detected by the  LSA (top panel) and PDM (middle panel) algorithms. The LSA minimum FAP is at half the correct period. The phased lightcurve (using the final refined period) of the star (V-728) is shown in the bottom panel.}
\label{fig:pdmsample}
\end{figure}

The Phase Dispersion Minimization (PDM) method~\citep{Stellingwerf1978} was also used to refine the periods of detected  variable stars. Our tests showed that this algorithm was more accurate than LSA in identifying the correct period for non-sinusoidal variables.  Figure~\ref{fig:pdmsample} shows the LSA and PDM periodograms for a sample eclipsing binary lightcurve. 

Variable star candidates were chosen from the complete data set by first taking all stars with LSA FAP below the adopted threshold. Candidate variables with FAP below an initial threshold had their periods refined by further application of LSA or PDM with a smaller trial period interval.
Visual examination of the phased lightcurves was used to reject many of these
initial candidates. The majority of rejected candidates had non-continuous phased lightcurves with periods close to 1 day.  A further systematic search for close variables with the same period led to the rejection of other candidates.

\begin{figure}
\resizebox{\hsize}{!}{\includegraphics{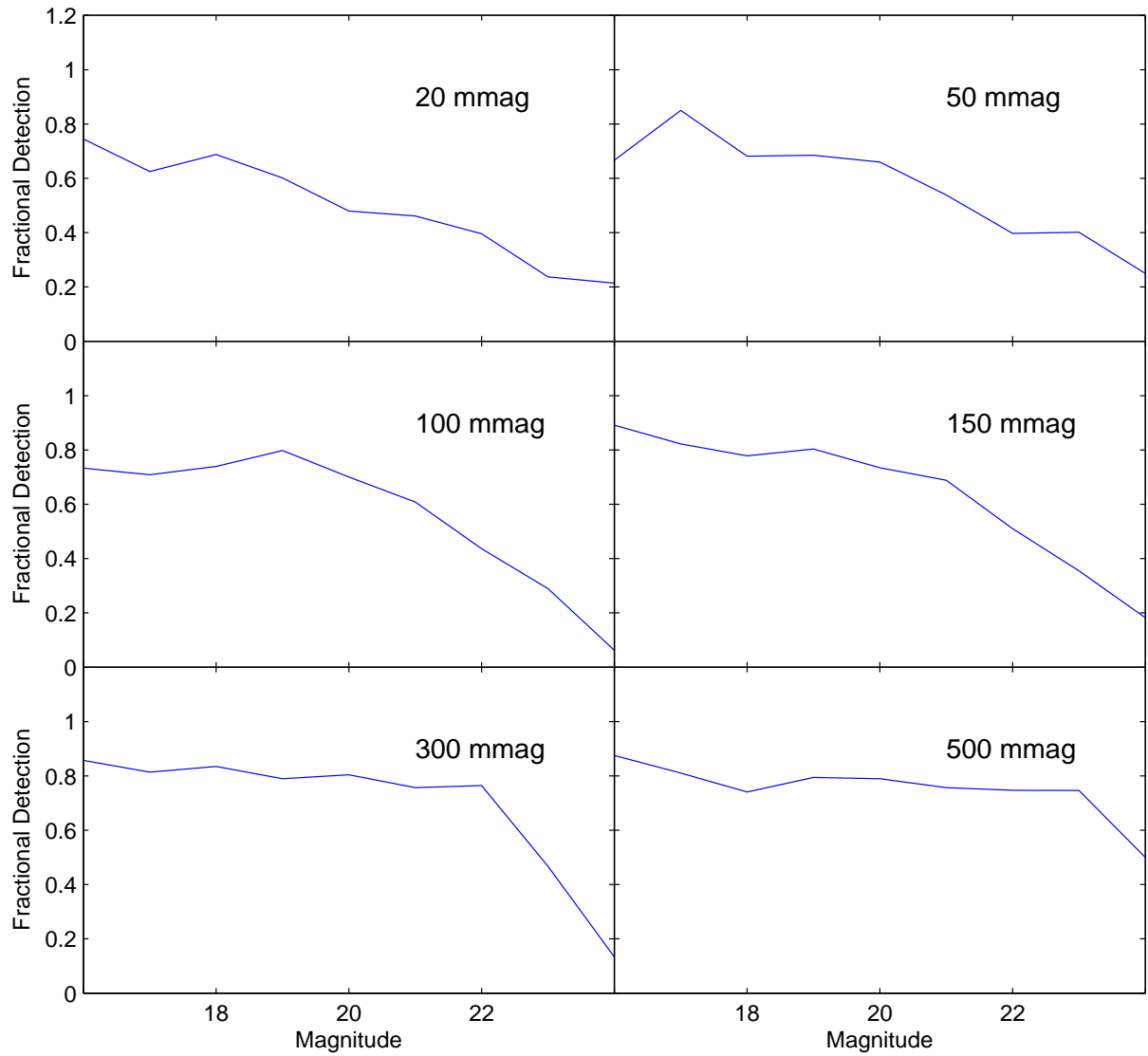}}
\caption{Results from insertion of simulated sinusoidal variables into real data and searched with the variable detection algorithm - panels show correctly detected variables as a fraction of inserted sinusoids as a function of R magnitude for each amplitude. 
}
\label{fig:vardeteff}
\end{figure}

By comparing the number of simulated sinusoidal variables with LSA FAP below the adopted threshold and 
recovered periods within 10\% of the inserted period or an alias, we have assessed our variable-star detection efficiency (Figure~\ref{fig:vardeteff}).  The efficiency peaks at 90\% for bright, high-amplitude variables. 
For lower amplitudes the detection efficiency drops significantly for fainter stars.

\section{Variable catalogue}                 \label{sec:gpvar}
A total of 1318 variable stars have been identified, only one of which is previously known (see \S\ref{sec:kv} for more details). The catalogue includes a large number of eclipsing binaries with 143 of Agol type, 335 of $\beta$ Lyrae type and 533 contact binaries (W Uma type). Also found are 9 possible Cepheids (DCEP) and a large number of un-categorized pulsating stars (PUL). 

The catalogue (Table~\ref{tab:vartable}) lists the star identification number, right ascension, declination, calibrated magnitudes of R, B-R and R-I where available, the period, variable type and any notes. A classification that is not certain is followed by a colon (i.e. EA:), a star of both types is given both designations (i.e. EA + PUL) and where there is ambiguity in the class the two types are combined with a slash (i.e. EA/PUL).
The notes column contains any other information ascertained about the star.

\subsection{Previously known variables}			\label{sec:kv}
Using the \textit{VizieR}~\citep{Ochsenbein2000} on-line database we queried the Combined General Catalogue of Variable Stars (GCVS)~\citep{Samus2004} which includes those stars labelled New Suspected Variable (NSV) and also the All Sky Automated Survey (ASAS)~\citep{Pojmanski2004} which has a catalogue of Southern Variables. A 30' cone search was performed at the field coordinates.

The search returned thirteen variable stars within the search area. These were examined individually by comparing our reference images to the DSS image of the region displayed by Aladin~\citep{Aladin} overlain with the variable stars found by \textit{VizieR}. Four of the thirteen fell outside the frame of the observations and three others appeared between the CCD chips. One set of coordinates belonged to an ASAS V band lightcurve and was not detected in our images. The remaining five were detected in our images but of the five, four were strongly saturated and not in our DAOPHOT catalogue. The final star (HO Nor, an Agol type eclipsing binary with a period of 2.12317 days) was saturated in most frames and our lightcurve contains only a few points. 

The remaining 1317 variables found from this survey are  previously unknown. 

\section{Contact binaries}		\label{sec:ewdis}

Contact binaries (EW) are amongst the most common variable stars. Contact binaries are detected at a rate of approximately 1 in 500 FGK dwarfs~\citep{Rucinski2006}. Here, we have catalogued 533 EW type stars out of 1318 variables.

Due to the similarities in lightcurve morphology, it is possible that some of  the stars we have categorized as EW are in fact BY Draconis stars. For such cases, the catalogued period is twice the rotation period. 

\begin{figure}
\resizebox{\hsize}{!}{\includegraphics{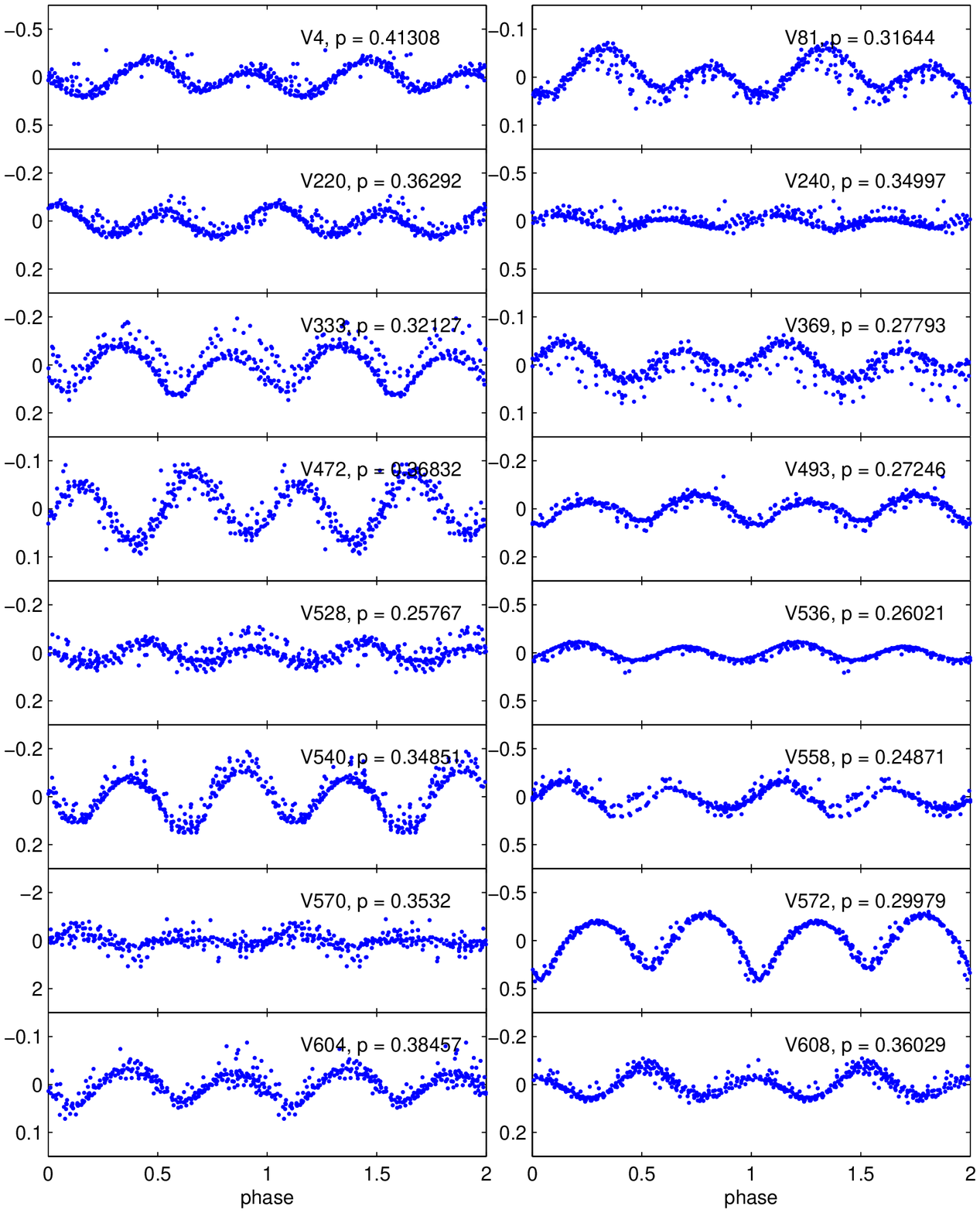}}
\caption{Lightcurves of contact binaries with maxima of different brightness.}
\label{fig:EW2max1}
\end{figure}

\begin{figure}
\resizebox{\hsize}{!}{\includegraphics{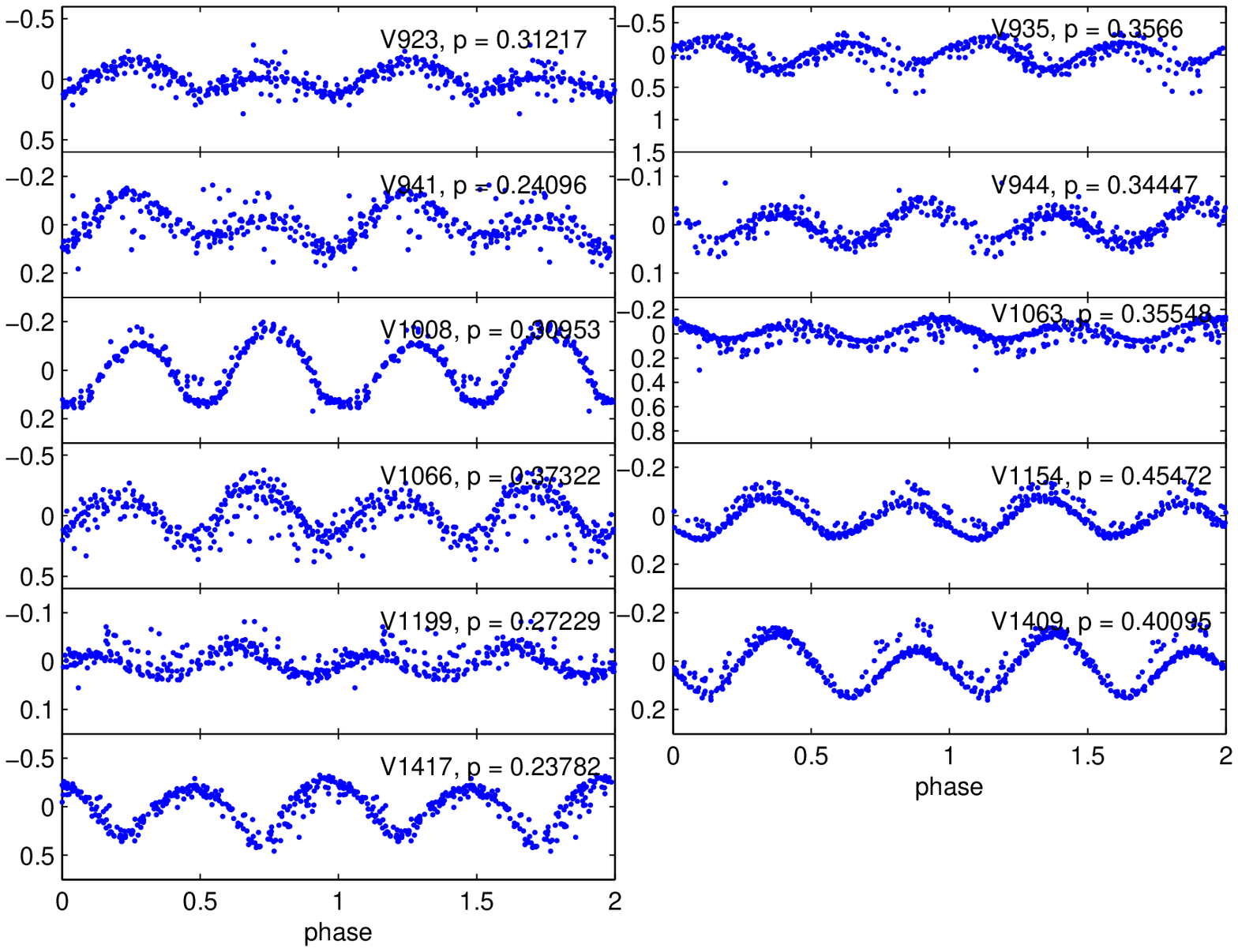}}
\caption{Lightcurves of contact binaries with maxima of different brightness.}
\label{fig:EW2max2}
\end{figure}

In the list of EW type stars there are a number with maxima at different brightness (the O'Connell effect)\citep{OConnell1951, Davidge1984}. These variables are listed in Table~\,\ref{tab:EW2max} with lightcurves in Figures~\,\ref{fig:EW2max1} and \,\ref{fig:EW2max2}.

Contact binaries have a strong period cut-off at approximately 0.215 to 0.22 days with a population maximum at slightly longer periods - around 0.27 days. The period cutoff is well known, but the reasons are not understood. \cite{Rucinski2007} has made a study of contact binaries found in the All-Sky Automated Survey. The ASAS survey contains more than 3000 contact binaries but includes only one (083128+1953.1) with a period less than 0.22 d.
That star has the shortest period of a contact eclipsing binary (0.2178 d) known prior to this paper. 

\begin{figure}
\resizebox{\hsize}{!}{\includegraphics{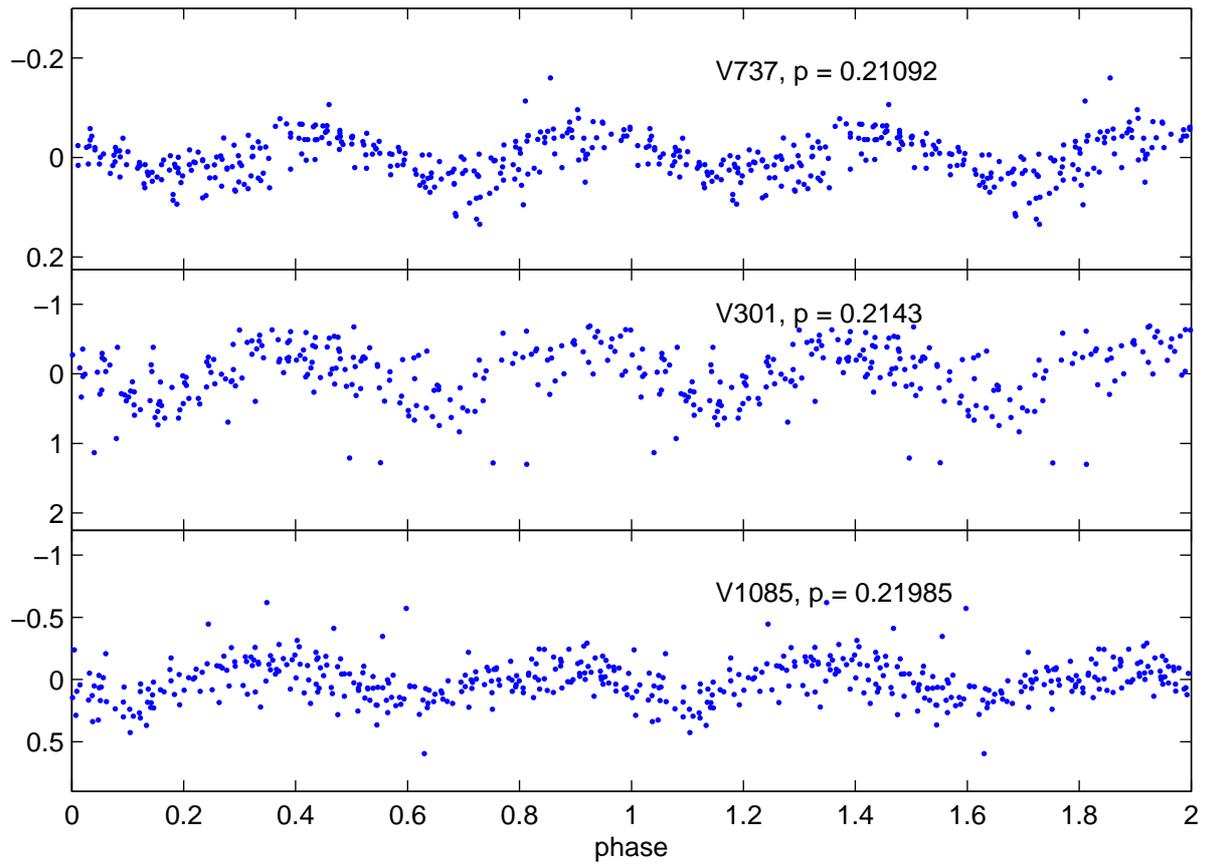}}
\caption{Lightcurves of the three probable contact binaries with periods of less than 0.22 days.}
\label{fig:EWPcut}
\end{figure}

From the variable stars found in this survey, we identified three as probable contact binaries with periods of less than 0.22 days, and another 4 with periods less than 0.23 days. 
Table~\ref{tab:EWPcut} lists the three shortest period candidates and their lightcurves are shown in Figure~\ref{fig:EWPcut}. All three systems are faint and it is possible that they are reddened pulsating stars, rather than binaries.  Followup photometry on the systems is desired to confirm their nature and verify the periods. Two of the stars have shorter periods than any contact binary published thus far.

\begin{figure}
\resizebox{\hsize}{!}{\includegraphics{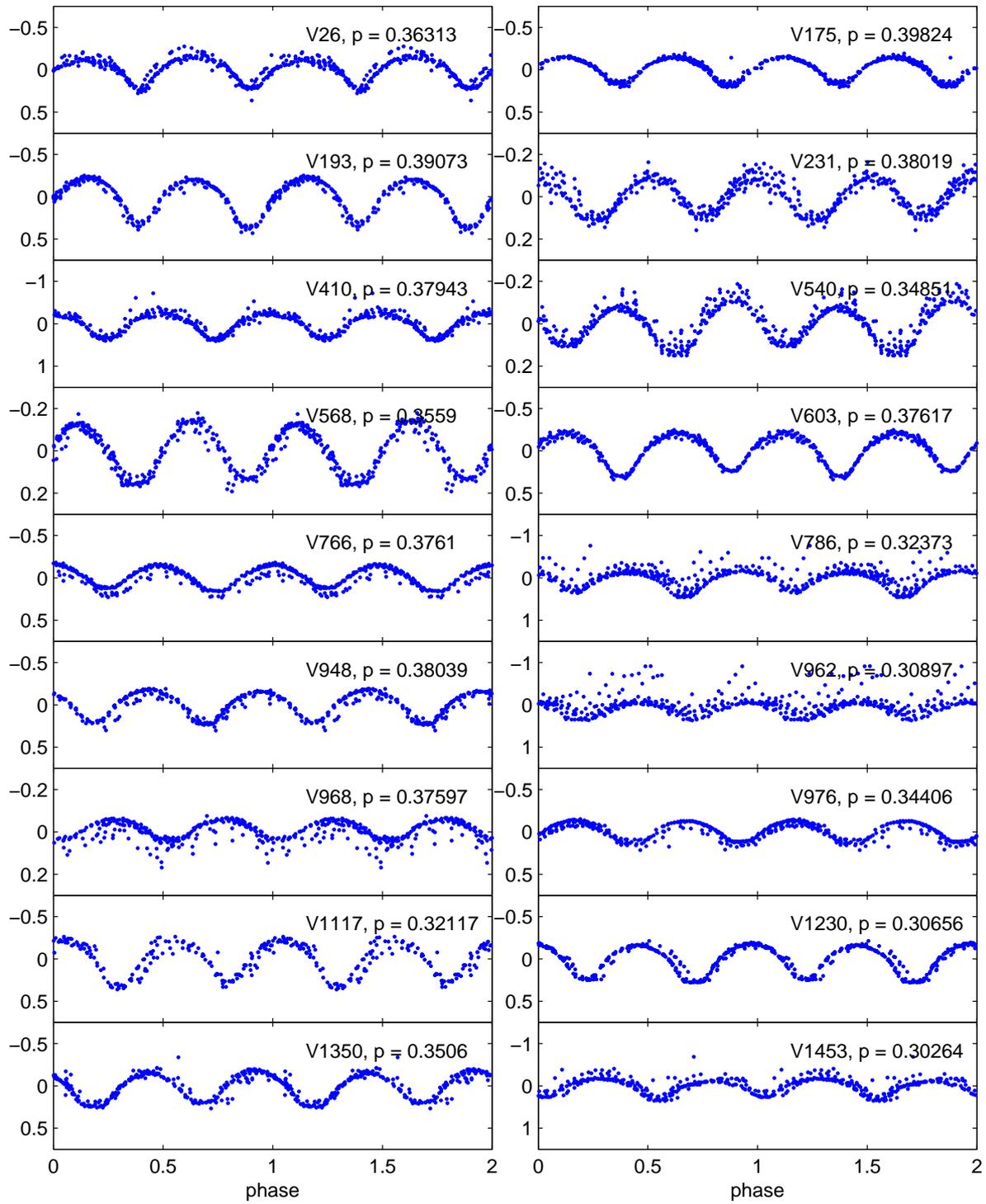}}
\caption{Lightcurves of the contact binaries with shape and periods suggesting low mass-ratios.}
\label{fig:EWlowmassratio}
\end{figure}

\begin{figure}
\resizebox{\hsize}{!}{\includegraphics{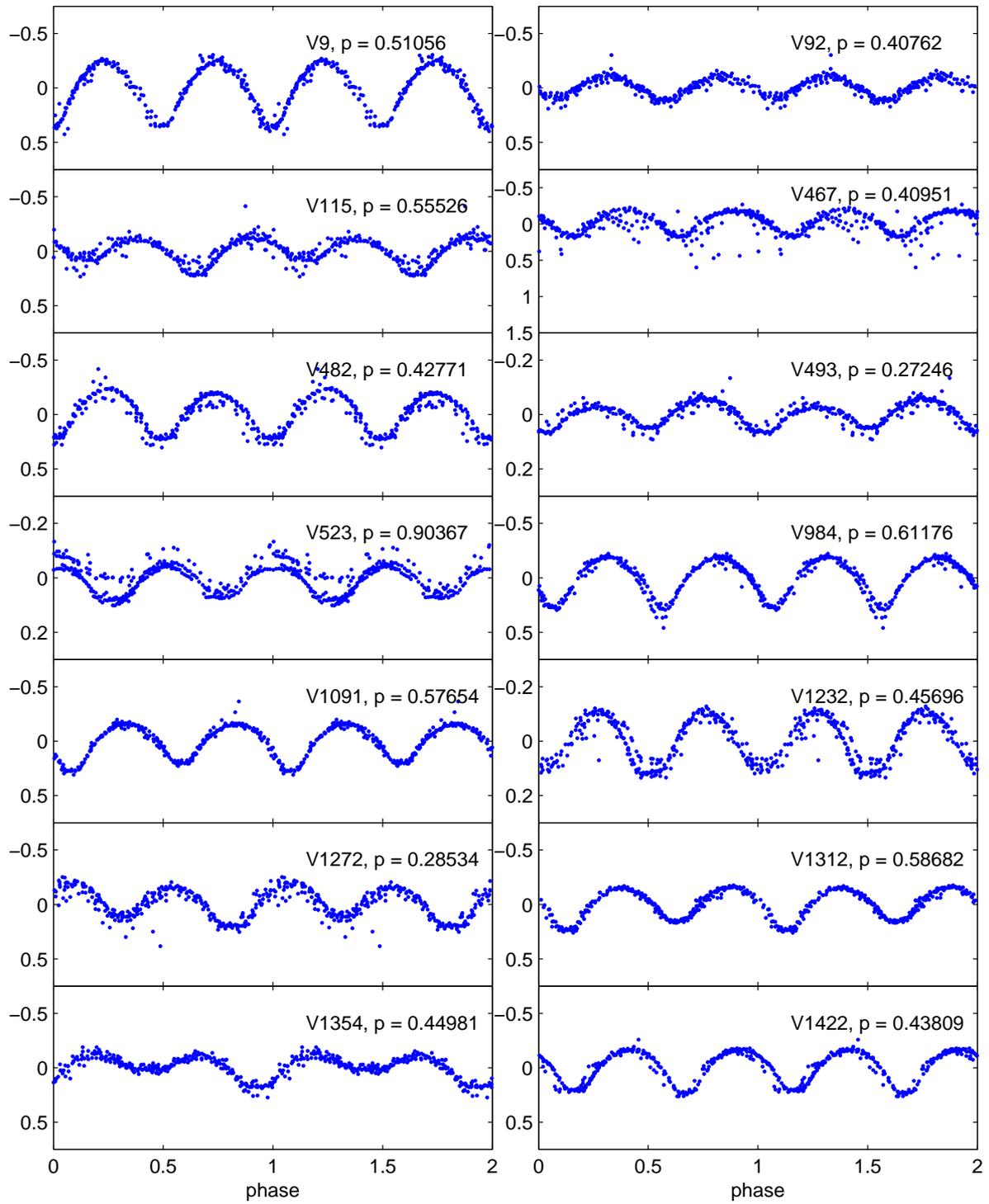}}
\caption{Lightcurves of the contact binaries with shape (but not period) suggesting low mass-ratios.}
\label{fig:EWlowmassratio2}
\end{figure}

The literature on contact binaries contains a number with low mass-ratios. The lightcurves of these systems always have a flat bottomed minimum and a period within the range 0.3 to 0.4 d. The flat bottom of the minimum is caused by the full eclipse of the smaller component of the binary. Our catalogue of contains 18 contact binaries whose lightcurves correspond to these criteria (see Table~\ref{tab:EWlowmass} and Figure~\ref{fig:EWlowmassratio}). There are also 14 contact binaries with flat bottomed lightcurves which fall outside the period range - 0.3 to 0.4d, (see Table~\ref{tab:EWlowmass2} and Figure~\ref{fig:EWlowmassratio2}).

\begin{figure}
\resizebox{\hsize}{!}{\includegraphics{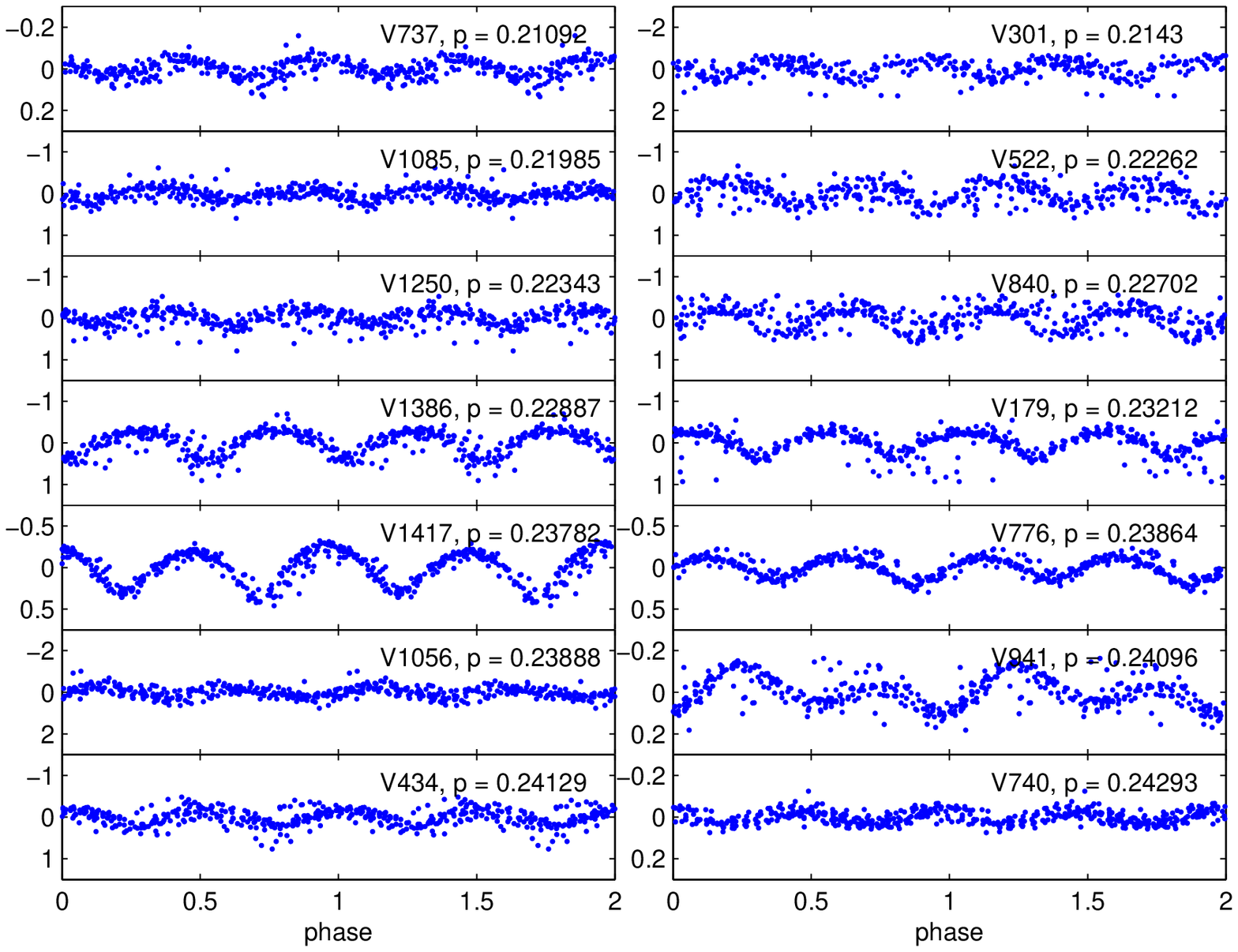}}
\caption{Lightcurves of eclipsing binaries with shape and periods suggesting a low-mass component.}
\label{fig:lowmasscomp1}
\end{figure}

\begin{figure}
\resizebox{\hsize}{!}{\includegraphics{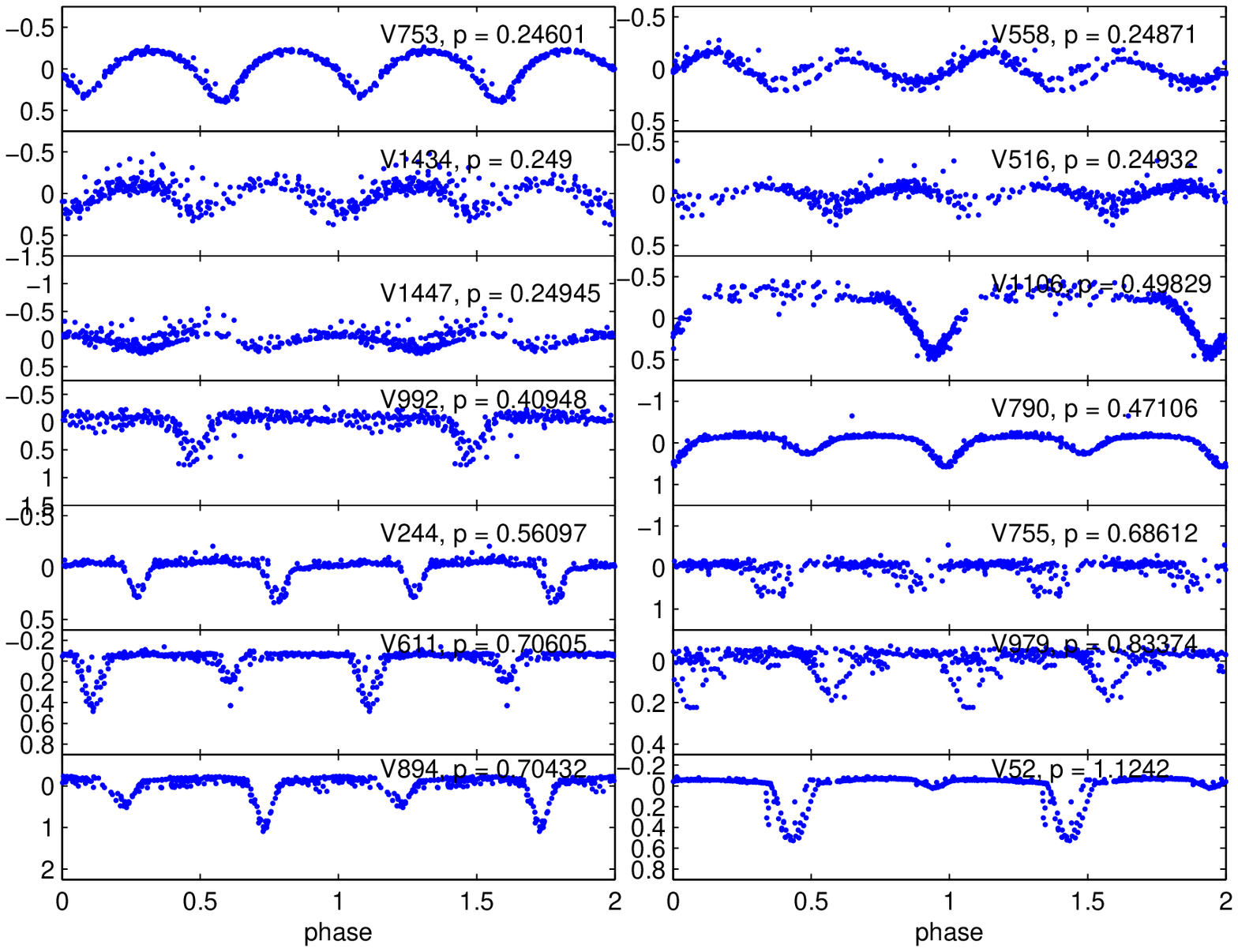}}
\caption{Lightcurves of the eclipsing binaries with shape and periods suggesting a low-mass component.}
\label{fig:lowmasscomp2}
\end{figure}

As for contact binaries with low mass-ratios it is possible to use the lightcurves found in the catalogue to identify systems that may have low-mass components. 
As in~\cite{Weldrake2007d}, we select contact binaries with periods of less than 0.25 days (19 candidates) and those Agol-type detached binaries with periods less than 1.6 days and non-varying out of eclipse lightcurves (9 candidates). The variables are listed in Tables~\ref{tab:lowmasscomp1} and~\ref{tab:lowmasscomp2} and their lightcurves shown in Figures~\ref{fig:lowmasscomp1} and \ref{fig:lowmasscomp2}. A number of these are brighter objects which would allow spectroscopic followup to confirm the nature of the components.

\subsection{{\bf Observed contact binary fraction}}			\label{sec:binfrac}
The observed fraction of contact binary stars in the field can be calculated from the number of detected contact binaries. 
As the variable star detection efficiency is not 100\% we must take this into account when we determine the binary fraction. We use the detection efficiency for an average amplitude (150 mmag) to determine an effective number of stars with detectible variations to be  62.6\% of the total.
From the fraction of detected contact eclipsing binaries we can make an estimate of the total fraction of contact binary stars in the field. There are 533 detected contact binaries. 
The effective number of field stars is 335592 giving an observed fraction of contact binaries as 0.16\%. Taking into account non-detections due to unfavourable orientations causes this number to double to 0.32\%, which tallies well with the rate of contact binaries found amongst main-sequence stars in the disk of the Galaxy by~\cite{Rucinski1997}.

\section{{\bf Other variable stars}}	

\subsection{Variable stars off the main sequence}		\label{sec:offmsvar}
Many variable stars can only be classified from photometric measurements by examining their position on a colour-magnitude diagram. The colour-magnitude diagrams for each subfield of our database are mostly comprised of main sequence stars, presumably the young population at the distance of the Norma spiral arm) plus a number of other field stars. There is considerable differential reddening between subfields. 

\begin{figure}
\resizebox{\hsize}{!}{\includegraphics{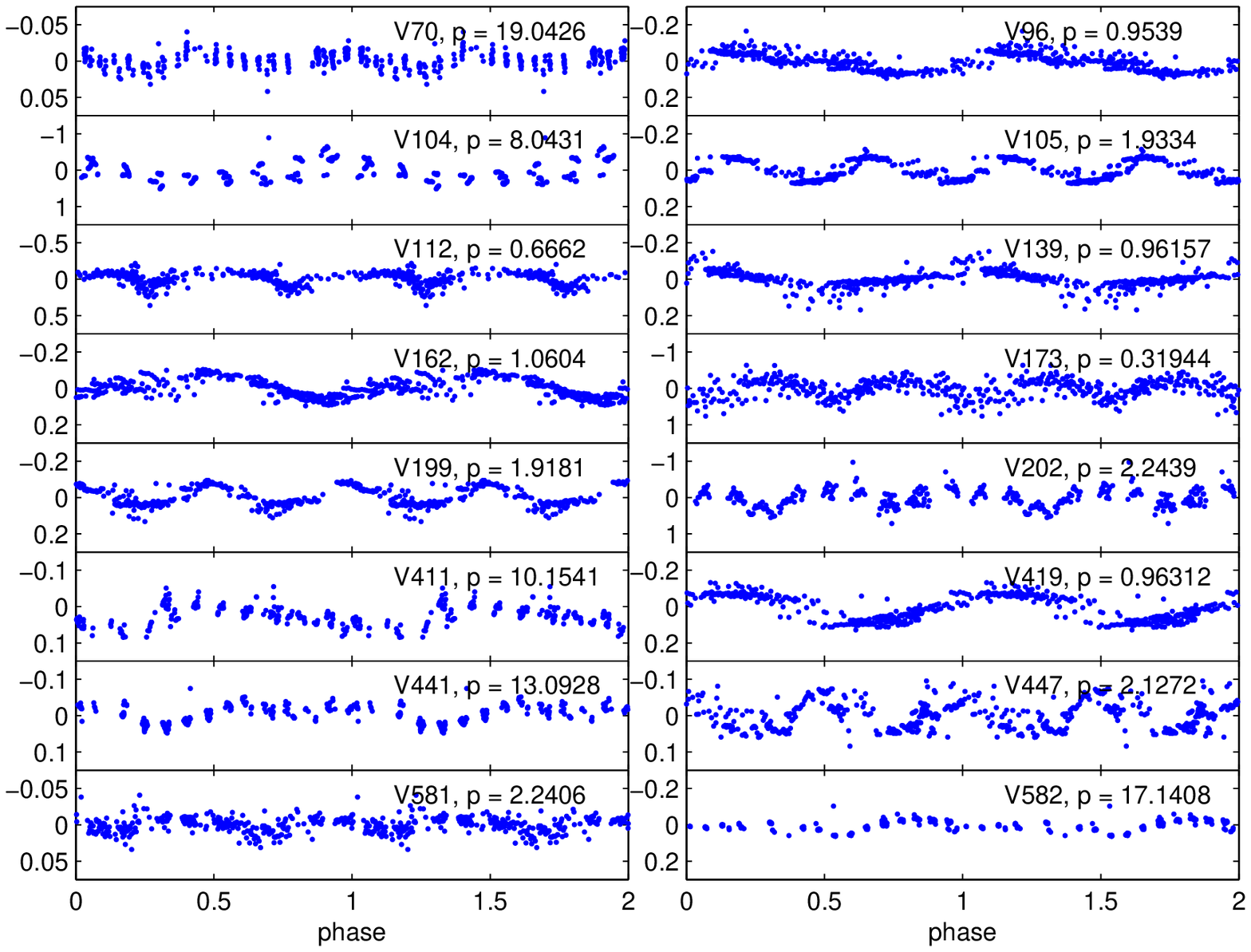}}
\caption{Lightcurves of variable stars which appear redder than the main sequence seen in the colour-magnitude diagrams.}
\label{fig:offms1}
\end{figure}

\begin{figure}
\resizebox{\hsize}{!}{\includegraphics{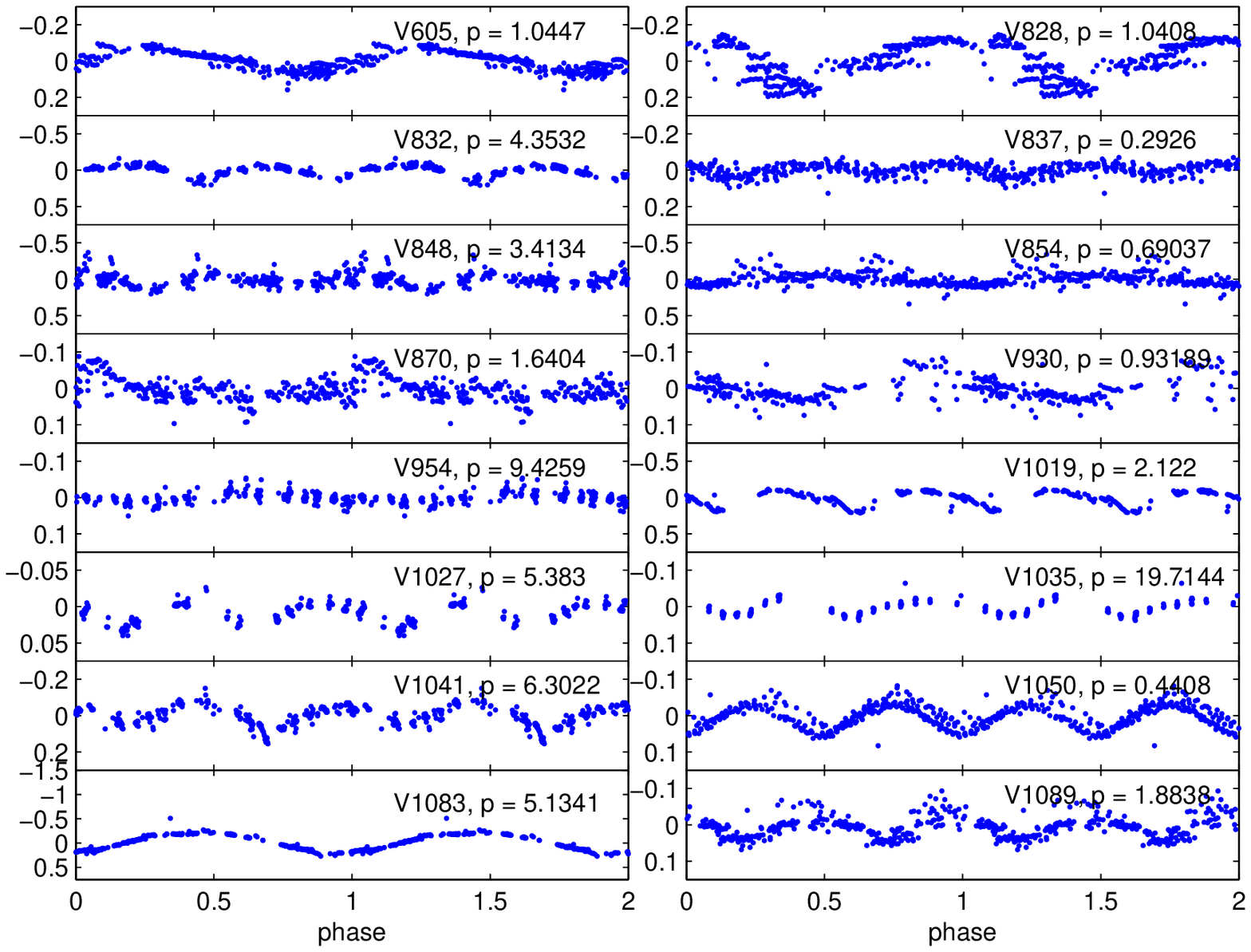}}
\caption{Lightcurves of variable stars which appear redder than the main sequence seen in the colour-magnitude diagrams.}
\label{fig:offms2}
\end{figure}

\begin{figure}
\resizebox{\hsize}{!}{\includegraphics{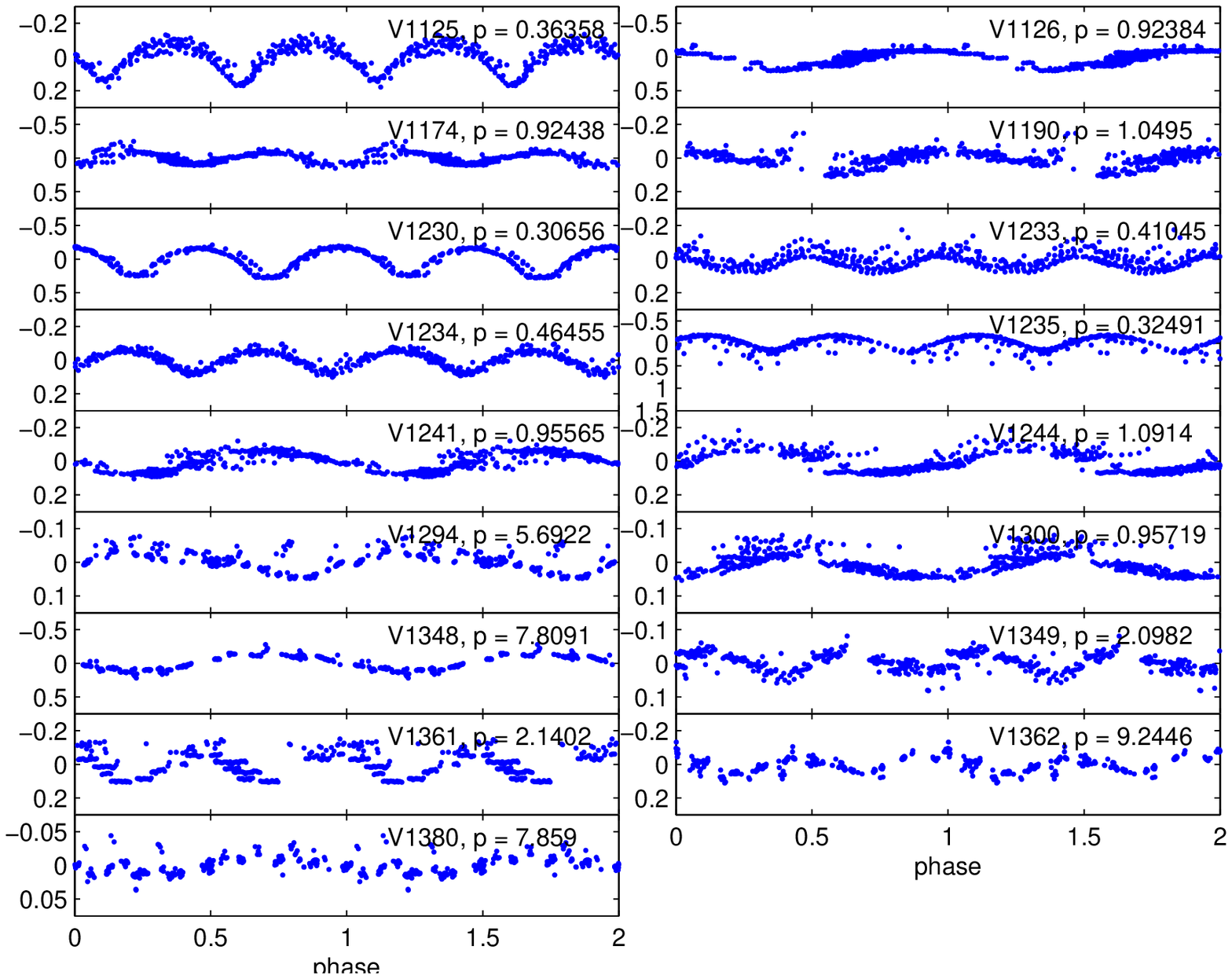}}
\caption{Lightcurves of variable stars which appear redder than the main sequence seen in the colour-magnitude diagrams.}
\label{fig:offms3}
\end{figure}

\begin{figure}
\resizebox{\hsize}{!}{\includegraphics{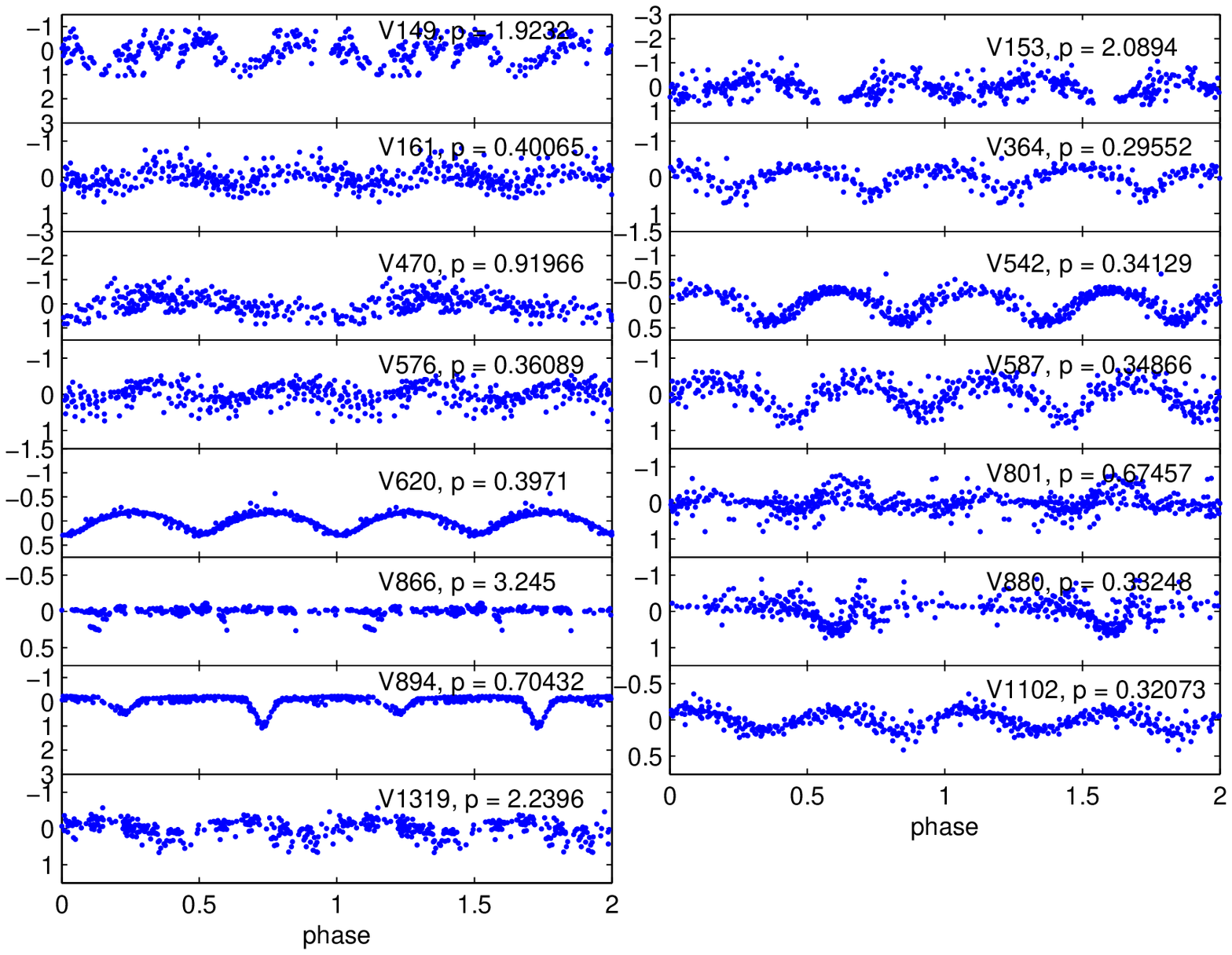}}
\caption{Lightcurves of variable stars which appear bluer than the main sequence seen in the colour-magnitude diagrams.}
\label{fig:offmsb}
\end{figure}

There are no features that can easily be interpreted as post-main sequence, although some of the stars may be evolved stars at greater distances. We have overplotted the positions of identified variable stars on the colour-magnitude diagrams for the appropriate subfields and found a number that appear to fall off the main sequence.
These stars form a disparate group with a large range of periods and variable types.
They are listed in Tables~\ref{tab:offmsvar} (redder than the main sequence) and ~\ref{tab:offmsvarb} (bluer). The lightcurves of these variables can be seen in Figures~\ref{fig:offms1} to~\ref{fig:offmsb}. 
The redder group includes the contact binary (V-1230) which  may have a low mass-ratio.
Binaries that are bluer than the main sequence may contain a white-dwarf component. A number of the bluer variables are eclipsing or ellipsoidal binaries and one of these stars (V-894) fits the criteria for a low mass companion.

\subsection{Pre-main sequence stars}				\label{sec:pms}
Detached eclipsing binaries are important for constraining stellar evolutionary theories. From detached eclipsing binaries one can determine elementary stellar information such as the mass and radius. Calibration of pre-main sequence (PMS) stars is scarce below 1 solar mass. There are only 6 PMS detached eclipsing binaries known in this mass range~\citep{Irwin2007b}. Fitting of these known PMS binaries with current models fails when both components are fitted simultaneously, implying a deficiency in the theoretical models. Detections of more pre-main sequence detached binaries will lead to stronger constraints on the theory. Following the method used in ~\cite{Christiansen2008b}, we use the positions of known pre-main sequence binaries on a colour-colour diagram, to determine which detached binaries in a sample may be pre-main sequence. These can then be followed up by spectroscopy~\citep{Irwin2007b}. In order for the comparison to be made, J,H and K magnitudes have been obtained from the 2MASS survey~\citep{2Mass} for the detached binaries in this survey so that they can be compared to the known pre-main sequence binaries.

\begin{figure}
\resizebox{\hsize}{!}{\includegraphics{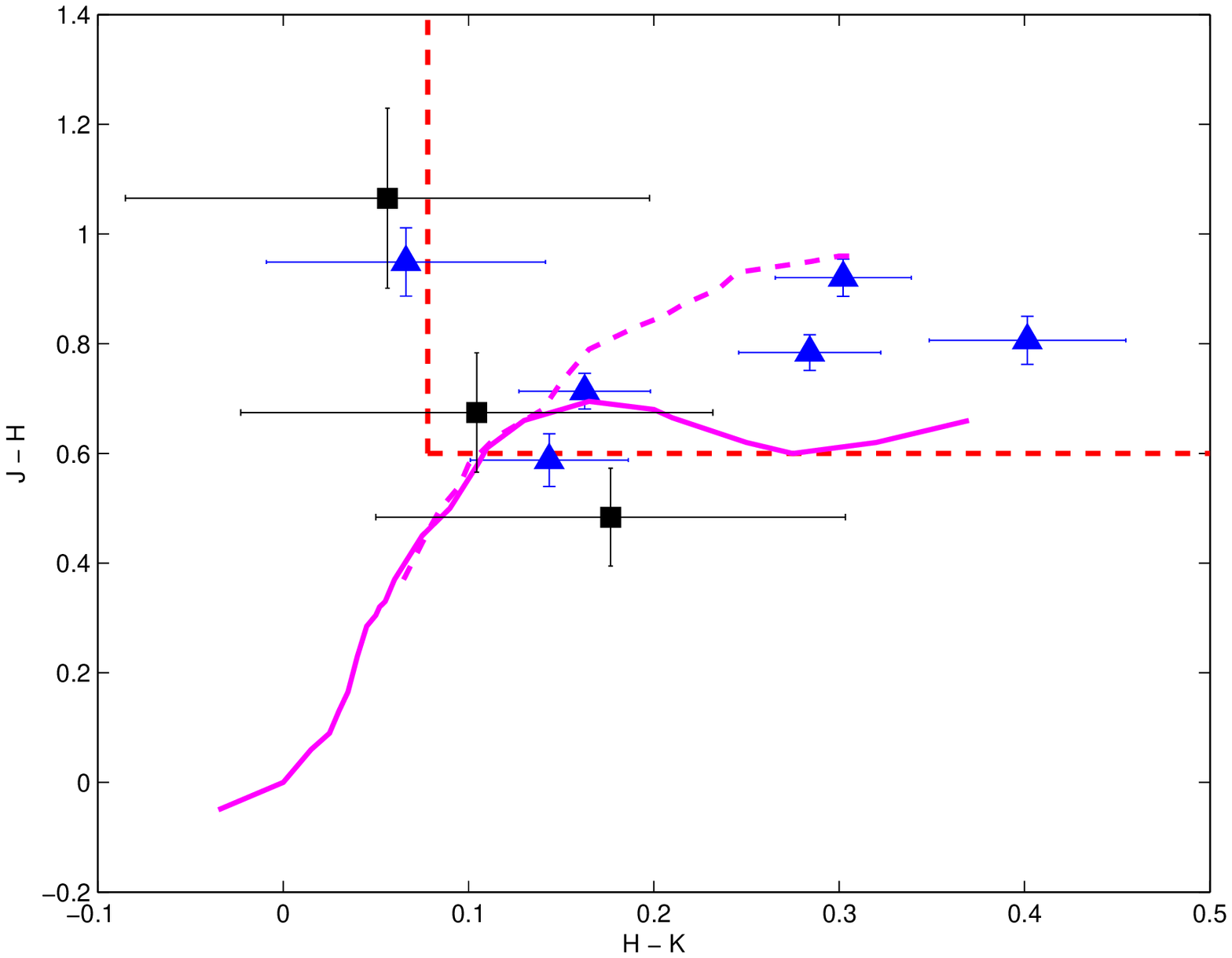}}
\caption{Colour-colour diagram in the infrared showing those known pre-main sequence eclipsing binaries (blue triangles) and candidates from this survey (black squares). The magnitudes of the known pre-main sequence stars are taken from \cite{Cargile2008,Covino2004,Hebb2006,Irwin2007b,Stassun2004} and \cite{Stassun2007}. The dashed lines represent limits on candidature of pre-main sequence binaries (J - H = 0.6 mag, H - K = 0.078 mag) Also plotted are the intrinsic stellar loci of giant stars (dashed line) and dwarf stars (solid line) from~\cite{Bessell1988}.}
\label{fig:PMS}
\end{figure}

Due to the faintness of many of the stars in the field only 3 of the EA type stars (V-827, V-887 and V-897) were given magnitudes by the 2MASS survey. A colour-colour diagram of the 6 known PMS eclipsing binaries and our three EA stars with 2MASS magnitudes is plotted in Figure~\ref{fig:PMS}. The dashed lines represent limits on candidature of pre-main sequence binaries (J - H = 0.6 mag, H - K = 0.078 mag) Also plotted are the intrinsic stellar loci of giant stars and dwarf stars  from~\cite{Bessell1988}.  The 2MASS measurements have been transformed into the Bessell \& Brett system using the equations in \citet{Carpenter2001}. All three of  our EA stars with 2MASS measurements are candidates for pre-main sequence stars (Table~\ref{tab:pms}) and worthy of further study.

\subsection{Other variable stars}			
\begin{figure}
\resizebox{\hsize}{!}{\includegraphics{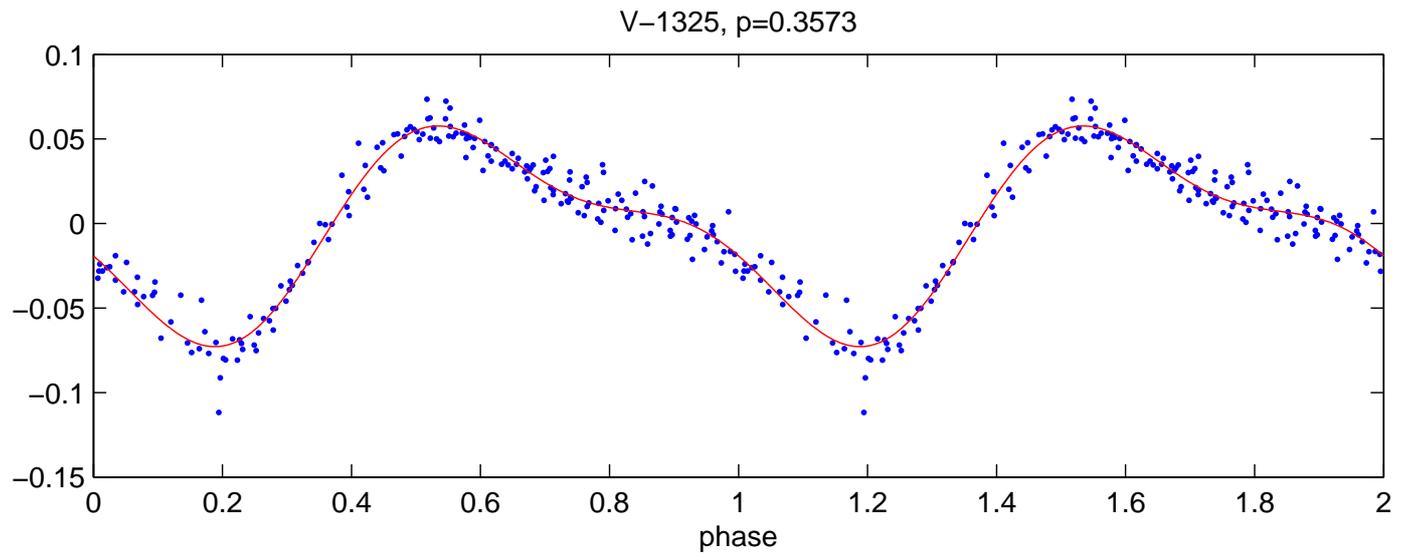}}
\caption{Lightcurve of a probable RR Lyrae star, shown with a fit that includes fundamental and first-harmonic components.}
\label{fig:RRL1}
\end{figure}

In cataloguing the variable stars a number of stars have been identified as miscellaneous pulsating stars. These stars are un-categorized either because the lightcurve showed no particular morphology that could be clearly identified or secondary information that was not available was needed. A number of these stars had lightcurves suggesting RR Lyrae type stars but further investigation of their position on the colour-magnitude diagram found them squarely on the main sequence. An example of such a star is shown in Figure~\ref{fig:RRL1}.

\section{Summary}		

From a photometric imaging survey of a 0.25 square degree region in Norma, we have extracted $\sim$500,000 lightcurves. 1318 variable stars have been identified, the majority of which are binaries.
A number of these stars are of interest for further study.
Photometric observations of the short-period contact binary stars would be useful to confirm the period and nature of these stars. Also needing followup are the low mass-ratio contact binaries to confirm their nature.
Those bright enough low-mass component candidates require spectroscopic measurements to confirm their nature. There are a number of Agol-type eclipsing binaries that may, given deeper infra-red measurements, be recognised as pre-main sequence binaries. 
Already there are  three candidate pre-main sequence stars (with infra-red observations from 2MASS) which require spectroscopic followup.

\section{Acknowledgements}
We thank the anonymous referee for his thoughtful comments. 
This work was supported by the Marsden Fund of New Zealand under contract UOC302.

\clearpage

\clearpage

\clearpage

\clearpage

\clearpage

\clearpage

\clearpage

\clearpage

\clearpage

\clearpage

\clearpage

\clearpage

\clearpage

\clearpage

\clearpage

\clearpage

\clearpage

\clearpage

\begin{table}
\caption{Summary of the observations used in this survey}
\label{tab:totobs}
\centering
\begin{tabular}{lllll}
\hline\hline
Year & Date & Filter & Number & Exposure time \\
\hline
2002 & 25 Jun - 15 Aug & R & 63 & 600  \\
2005 & 26 Jul - 25 Aug & R & 204 & 720 \\
2005 & 26 Jul - 25 Aug & B & 3 & 720 \\
2005 & 26 Jul - 25 Aug & I & 3 & 720 \\
\hline
\end  {tabular}
\end {table}

\clearpage

\clearpage

\begin{longtable}{llllllllll}
\caption{\label{tab:vartable}Catalogue of variable stars.Tabulated are star ID, coordinates, magnitudes, period and variable type}\\
\hline\hline
ID & Subset & RA & DEC & R & B-R & R-I & Period & Type & Notes\\
& & & & & & & (days) & & \\
\hline
\endfirsthead
\caption{continued.}\\
\hline\hline
ID & Subset & RA & DEC & R & B-R & R-I & Period & Type & Notes\\
& & & & & & & (days) & & \\
\hline
\endhead
\hline
\endfoot
V-1 & 1A & 16:20:37.72	&-53:25:26.5011 & 17.84 & 1.81 & 1.90 & 2.18480 & EB & \\			
V-2 & 1A & 16:20:52.98	&-53:25:24.0178 & 19.68 & 1.96 & 1.71 & 0.43817 & EW & \\			
V-3 & 1A & 16:20:41.89	&-53:25:07.5671 & 20.27 & 2.30 & 2.23 & 0.67577 & EW & \\			
V-4 & 1A & 16:20:47.06	&-53:25:06.8219 & 18.66 & 1.67 & 2.05 & 0.41308 & EW & 2	max	\\	
V-6 & 1A & 16:20:49.68	&-53:24:54.9648 & 15.97 & 1.26 & 1.64 & 8.45680 & PUL: & \\			
V-7 & 1A & 16:20:41.60	&-53:24:52.8653 & 17.94 & 1.58 & 1.76 & 1.87580 & EB & \\			
V-9 & 1A & 16:20:50.50	&-53:24:40.1987 & 19.05 & 1.33 & 2.28 & 0.51056 & EW & flat	bottom	\\	
V-11 & 1A & 16:20:46.81	&-53:24:35.0235 & 19.31 & 2.16 & 1.82 & 0.40644 & EW & \\
\end{longtable}

\begin{table*}
\centering
\caption
[Contact binaries with two maxima]
{Table of contact binaries with maxima of different brightness.}
\label{tab:EW2max}
\begin{tabular}{llllllllll}
\hline\hline
ID & Subset & Right Ascension & Declination & \multicolumn{3}{c}{Calibrated Magnitudes} & Period & Type \\ 
& &(J2000.0) & (J2000.0) & R & B-R & R-I & (days) &\\
\hline
V-4 & 1A & 16:20:47.06 &-53:25:06.8219 & 18.66 & 1.67 & 2.05 & 0.4131 & EW \\
V-81 & 1C & 16:20:34.42 &-53:20:30.1431 & 17.77 & 1.95 & 0.68 & 0.3164 & EW \\
V-220 & 2B & 16:19:26.09 &-53:23:07.7984 & 17.35 & 2.00 & 0.43 & 0.3629 & EW \\
V-240 & 2C & 16:19:39.44 &-53:19:3.5470 & 18.12 & 2.34 & 0.59 & 0.3500 & EW \\
V-333 & 2G & 16:19:37.64 &-53:11:39.8602 & 18.06 & 2.94 & 0.43 & 0.3213 & EW \\
V-369 & 3A & 16:18:59.77 &-53:25:09.8420 & 16.74 & 2.00 & 0.19 & 0.2779 & EW \\
\hline
\end{tabular}
\end{table*}

\clearpage

\begin{table*}
\caption{Contact binaries with periods close to that of the known period cutoff.}
\label{tab:EWPcut}
\centering
\begin{tabular}{lllllllll}
\hline\hline
ID & Subset & Right Ascension & Declination & \multicolumn{3}{c}{Calibrated Magnitudes} & Period & Type \\ 
& & (J2000.0) & (J2000.0) & R & B-R & R-I & (days) & \\
\hline
V-737 & 4G & 16:17:50.67 &-53:13:47.9457 & 19.19 & 2.65 & 1.16 & 0.2109 & EW \\
V-301 & 2F & 16:19:27.64 &-53:16:24.3907 & 22.14 &  & 1.68 & 0.2143 & EW \\
V-1085 & 6F & 16:18:14.03 &-53:32:50.9721 & 21.21 & 2.80 & 1.33 & 0.2199 & EW \\
\hline
\end{tabular}
\end{table*}

\clearpage

\begin{table*}
\caption{Contact binaries that may have a low mass-ratio.}
\label{tab:EWlowmass}
\centering
\begin{tabular}{lllllllll}
\hline\hline
ID & Subset & Right Ascension & Declination & \multicolumn{3}{c}{Calibrated Magnitudes} & Period & Type \\ 
& &(J2000.0) & (J2000.0) & R & B-R & R-I & (days) & \\
\hline
V-26 & 1A & 16:20:43.91 &-53:23:29.7361 & 19.17 & 2.27 & 1.96 & 0.3631 & EW \\
V-175 & 1H & 16:20:08.57 &-53:11:51.7763 & 17.98 & 2.34 & 0.81 & 0.3982 & EW \\
V-193 & 2A & 16:19:38.23 &-53:24:59.8532 & 16.47 & 1.25 & 0.97 & 0.3907 & EW \\
V-231 & 2C & 16:19:42.23 &-53:20:23.1327 & 18.42 & 2.30 & 0.48 & 0.3802 & EW \\
V-410 & 3B & 16:18:22.56 &-53:24:10.8596 & 19.03 & 1.69 & 1.00 & 0.3794 & EW \\
V-540 & 3G & 16:18:50.54 &-53:13:15.8250 & 18.10 & 2.13 & 0.52 & 0.3485 & EW \\
V-568 & 3H & 16:18:15.50 &-53:12:36.5601 & 17.37 & 2.03 & 0.66 & 0.3559 & EW \\
V-603 & 4A & 16:18:04.46 &-53:24:25.2664 & 15.73 & 1.92 & 0.47 & 0.3762 & EW \\
V-766 & 4H & 16:17:31.19 &-53:13:13.1713 & 17.07 & 1.93 & 0.51 & 0.3761 & EW \\
V-786 & 4H & 16:17:23.08 &-53:10:55.8408 & 18.15 & 2.34 & 0.71 & 0.3237 & EW \\
V-948 & 5G & 16:18:04.49 &-53:27:07.6592 & 17.07 & 2.09 & 0.05 & 0.3804 & EW \\
V-962 & 5H & 16:17:27.09 &-53:28:59.5381 & 19.12 & 2.30 & 0.62 & 0.3090 & EW \\
V-968 & 5H & 16:17:38.22 &-53:27:28.1100 & 16.90 & 1.64 & 0.59 & 0.3760 & EW \\
V-976 & 6A & 16:19:00.14 &-53:41:37.2113 & 17.33 & 2.45 & 0.04 & 0.3441 & EW \\
V-1117 & 6H & 16:18:36.27 &-53:28:26.7849 & 19.44 & 2.35 & 0.56 & 0.3212 & EW \\
V-1230 & 7E & 16:19:56.70 &-53:32:24.1384 & 15.89 & 1.06 & 0.38 & 0.3066 & EW \\
V-1350 & 8C & 16:20:34.33 &-53:37:18.2836 & 18.30 & 2.13 & 0.56 & 0.3506 & EW \\
V-1453 & 8G & 16:20:45.64 &-53:27:00.6708 & 18.61 & 2.40 & 0.17 & 0.3026 & EW \\
\hline
\end{tabular}
\end{table*}

\clearpage

\begin{table*}
\caption{Contact binaries that may have a low mass-ratio - the lightcurves have shape but not periods suggesting low mass-ratios.}
\label{tab:EWlowmass2}
\centering
\begin{tabular}{lllllllll}
\hline\hline
ID & Subset & Right Ascension & Declination & \multicolumn{3}{c}{Calibrated Magnitudes} & Period & Type \\ 
& &(J2000.0) & (J2000.0) & R & B-R & R-I & (days) & \\
\hline
V-9 & 1A & 16:20:50.50 &-53:24:40.1987 & 19.05 & 1.33 & 2.28 & 0.5106 & EW \\
V-92 & 1C & 16:20:48.80 &-53:19:04.3178 & 19.24 & 2.23 & 0.94 & 0.4076 & EW \\
V-115 & 1E & 16:20:53.66 &-53:17:11.7310 & 19.14 & 2.55 & 0.67 & 0.5553 & EW \\
V-467 & 3E & 16:18:59.97 &-53:16:25.5253 & 19.25 & 2.83 & 0.52 & 0.4095 & EW \\
V-482 & 3E & 16:18:56.66 &-53:15:05.8831 & 17.15 & 1.95 & 0.53 & 0.4277 & EW \\
V-493 & 3F & 16:18:36.50 &-53:17:09.0202 & 16.23 & 1.87 & 1.75 & 0.2725 & EW \\
V-523 & 3F & 16:18:30.89 &-53:15:11.5286 & 16.49 & 2.38 &  & 0.9037 & EW \\
V-984 & 6A & 16:18:47.74 &-53:41:03.6335 & 17.74 & 2.84 & 0.61 & 0.6118 & EW \\
V-1091 & 6F & 16:18:38.73 &-53:31:56.7975 & 17.24 & 2.20 & 1.08 & 0.5765 & EW \\
V-1232 & 7E & 16:19:39.57 &-53:32:12.9626 & 18.13 & 1.65 & 0.90 & 0.4570 & EW \\
V-1272 & 7G & 16:19:38.92 &-53:27:04.6993 & 18.47 & 2.33 & 0.82 & 0.2853 & EW \\
V-1312 & 8A & 16:20:38.43 &-53:38:56.5849 & 17.00 & 1.24 & 4.28 & 0.5868 & EW \\
V-1354 & 8C & 16:20:48.55 &-53:36:20.9116 & 18.64 & 1.86 & 0.80 & 0.4498 & EW \\
V-1422 & 8F & 16:20:09.11 &-53:32:17.6392 & 18.48 & 1.92 & 0.75 & 0.4381 & EW \\
\hline
\end{tabular}
\end{table*}

\clearpage

\begin{table*}
\caption{Eclipsing binaries that may have a low mass component.}
\label{tab:lowmasscomp1}
\centering
\begin{tabular}{lllllllll}
\hline\hline
ID & Subset & Right Ascension & Declination & \multicolumn{3}{c}{Calibrated Magnitudes} & Period & Type \\ 
& &(J2000.0) & (J2000.0) & R & B-R & R-I & (days) & \\
\hline
V-737 & 4G & 16:17:50.67 &-53:13:47.9457 & 19.19 & 2.65 & 1.16 & 0.2109 & EW \\
V-301 & 2F & 16:19:27.64 &-53:16:24.3907 & 22.14 &  & 1.68 & 0.2143 & EW \\
V-1085 & 6F & 16:18:14.03 &-53:32:50.9721 & 21.21 & 2.80 & 1.33 & 0.2199 & EW \\
V-522 & 3F & 16:18:28.65 &-53:15:13.7680 & 21.54 & 3.46 & 2.35 & 0.2226 & EW \\
V-1250 & 7G & 16:19:51.13 &-53:29:59.1415 & 21.06 & 2.62 & 1.65 & 0.2234 & EW \\
V-840 & 5C & 16:17:57.44 &-53:36:59.9529 & 21.49 & 3.02 & 1.68 & 0.2270 & EW \\
V-1386 & 8D & 16:20:07.71 &-53:35:31.1146 & 20.01 & 1.91 & 2.30 & 0.2289 & EW \\
V-179 & 1H & 16:20:19.37 &-53:11:35.1909 & 20.51 & 3.03 & 1.13 & 0.2321 & EW \\
V-1417 & 8F & 16:20:31.49 &-53:33:04.9446 & 19.53 & 1.54 & 1.08 & 0.2378 & EW \\
V-776 & 4H & 16:17:34.08 &-53:12:02.1376 & 19.56 & 2.53 & 0.97 & 0.2386 & EW \\
V-1056 & 6D & 16:18:15.23 &-53:35:25.1012 & 21.87 & 2.96 & 1.21 & 0.2389 & EW \\
V-941 & 5G & 16:17:53.57 &-53:27:48.4901 & 18.66 & 2.58 & 0.84 & 0.2410 & EW \\
V-434 & 3C & 16:18:54.81 &-53:20:48.4850 & 20.18 & 2.62 & 0.66 & 0.2413 & EW \\
V-740 & 4G & 16:17:49.64 &-53:13:21.5601 & 17.91 & 2.34 & 0.96 & 0.2429 & EW \\
V-753 & 4G & 16:17:56.34 &-53:11:51.1016 & 18.01 & 2.20 & 0.40 & 0.2460 & EW \\
V-558 & 3G & 16:18:54.13 &-53:11:08.7186 & 18.95 & 2.33 & 0.85 & 0.2487 & EW \\
V-1434 & 8F & 16:20:20.06 &-53:31:01.2833 & 20.47 & 2.48 & 0.89 & 0.2490 & EW \\
V-516 & 3F & 16:18:31.63 &-53:15:35.7031 & 19.64 & 3.42 & 2.40 & 0.2493 & EW \\
V-1447 & 8G & 16:20:51.11 &-53:27:48.4544 & 17.43 & 2.31 & 0.40 & 0.2494 & EW \\
\hline
\end{tabular}
\end{table*}

\clearpage

\begin{table*}
\caption{Detached eclipsing binaries that may have a low mass component.}
\label{tab:lowmasscomp2}
\centering
\begin{tabular}{lllllllll}
\hline\hline
ID & Subset & Right Ascension & Declination & \multicolumn{3}{c}{Calibrated Magnitudes} & Period & Type \\ 
& &(J2000.0) & (J2000.0) & R & B-R & R-I & (days) & \\
\hline
V-1106 & 6G & 16:18:55.35 &-53:28:34.7387 & 18.83 & 2.19 & 0.73 & 0.4983 & EA \\
V-992 & 6A & 16:18:45.09 &-53:39:45.9131 & 19.46 & 2.46 & 1.48 & 0.4095 & EA \\
V-790 & 4H & 16:17:28.84 &-53:10:42.0607 & 17.83 & 2.09 & 0.37 & 0.4711 & EA \\
V-244 & 2C & 16:19:37.54 &-53:18:34.5958 & 18.35 & 2.84 & 1.14 & 0.5610 & EA \\
V-755 & 4G & 16:17:58.34 &-53:11:39.78 & 20.63 & 3.00 & 1.28 & 0.6861 & EA \\
V-611 & 4A & 16:18:02.91 &-53:23:07.2723 & 18.33 & 2.37 & 0.50 & 0.7061 & EA \\
V-979 & 6A & 16:18:46.94 &-53:41:25.8984 & 18.57 & 2.91 & 0.43 & 0.8337 & EA \\
V-894 & 5E & 16:18:05.07 &-53:31:34.5940 & 19.24 & 1.60 & 1.65 & 0.7043 & EA \\
V-52 & 1B & 16:20:11.65 &-53:24:02.0859 & 16.28 & 1.86 & 0.19 & 1.1242 & EA \\
\hline
\end{tabular}
\end{table*}

\clearpage

\begin{table*}
\caption{Variable stars redder than the main sequence.}
\label{tab:offmsvar}
\centering
\begin{tabular}{lllllllll}
\hline\hline
ID & Subset & Right Ascension & Declination & \multicolumn{3}{c}{Calibrated Magnitudes} & Period & Type \\ 
& &(J2000.0) & (J2000.0) & B & I & R & (days) &\\
\cr\hline
V-70 & 1C & 16:20:42.84 &-53:21:34.5504 & 17.13 & 2.76 & 0.97 & 19.0426 & EB \\
V-96 & 1C & 16:20:51.92 &-53:18:49.0723 & 17.95 & 2.99 & 1.01 & 0.9539 & PUL: \\
V-104 & 1D & 16:20:14.62 &-53:21:05.4916 & 17.58 & 2.11 & 0.75 & 8.0431 & PUL: \\
V-105 & 1D & 16:20:12.64 &-53:20:57.3738 & 16.27 & 3.34 &  & 1.9334 & EB/PUL \\
V-112 & 1D & 16:20:19.04 &-53:19:30.1081 & 20.02 & 3.23 & 1.63 & 0.6662 & EW \\
\hline
\end{tabular}
\end{table*}

\clearpage

\begin{table*}
\caption{Variable stars bluer than the main sequence}
\label{tab:offmsvarb}
\begin{tabular}{lllllllll}
\hline\hline
ID & Subset & Right Ascension & Declination & \multicolumn{3}{c}{Calibrated Magnitudes} & Period & Type \\ 
& &(J2000.0) & (J2000.0) & B & I & R & (days) &\\
\hline
V-149 & 1G & 16:20:56.01 &-53:11:41.6310 & 21.46 & 1.11 & 1.18 & 1.9232 & EB: \\
V-153 & 1G & 16:20:48.70 &-53:11:24.4095 & 22.08 &  & 1.18 & 2.0894 & EB: \\
V-161 & 1G & 16:20:38.50 &-53:10:42.5178 & 21.33 & 2.35 & 0.99 & 0.4007 & EW: \\
V-364 & 3A & 16:19:01.65 &-53:25:40.0235 & 21.38 & 2.24 & 1.29 & 0.2955 & EW \\
V-470 & 3E & 16:18:41.70 &-53:16:22.4156 & 21.87 & 1.68 & 0.27 & 0.9197 & PUL: \\
V-542 & 3G & 16:18:52.29 &-53:13:13.8844 & 20.12 & 1.82 & 0.69 & 0.3413 & EW \\
V-576 & 3H & 16:18:18.61 &-53:12:13.2789 & 21.19 & 2.05 & 0.91 & 0.3609 & EW \\ 
V-587 & 3H & 16:18:21.26 &-53:10:49.3220 & 21.55 & 2.32 & 1.32 & 0.3487 & EW \\
V-620 & 4B & 16:17:36.00 &-53:25:04.7264 & 18.35 & 1.81 & 0.63 & 0.3971 & EW \\
V-801 & 5A & 16:18:01.48 &-53:40:19.5726 & 20.59 & 1.83 & 0.34 & 0.6746 & CV/PUL \\
V-866 & 5D & 16:17:17.11 &-53:35:29.5915 & 18.43 & 2.18 & 0.34 & 3.2450 & EA: \\
V-880 & 5E & 16:17:48.20 &-53:32:53.2065 & 19.03 & 1.41 & 1.24 & 0.3325 & EW/EA \\
V-894 & 5E & 16:18:05.07 &-53:31:34.5940 & 19.24 & 1.60 & 1.65 & 0.7043 & EA \\
V-1102 & 6G & 16:18:43.98 &-53:29:36.7935 & 19.92 & 0.82 & 1.07 & 0.3207 & EW \\
V-1319 & 8B & 16:20:20.87 &-53:41:30.5543 & 21.63 & 0.70 & 1.79 & 2.2396 & EB: \\
\hline
\end{tabular}
\end{table*}

\clearpage

\begin{table*}
\caption{Possible pre-main sequence eclipsing binaries} 
\label{tab:pms}
\centering
\begin{tabular}{llllllllll}
\hline\hline
ID & Subset & 2MASS ID & Right Ascension & Declination & \multicolumn{3}{c}{Calibrated Magnitudes} & Period & Type \\ 
& & &(J2000.0) & (J2000.0) & R & B-R & R-I & (days) & \\
\hline
V-827 & 5B & 16173776-5339210 & 16:17:37.78 &-53:39:20.1204 & 16.20 & 2.10 & 0.31 & 0.5727 & EA \\
V-887 & 5E & 16174598-5332177 & 16:17:45.99 &-53:32:17.4920 & 16.27 & 1.60 & 1.01 & 2.2913 & EA \\
V-897 & 5E & 16175274-5331011 & 16:17:52.77 &-53:31:02.2072 & 15.89 & 1.75 & 1.07 & 7.59400 & EA \\	
\hline
\end{tabular}
\end{table*}

\end{document}